\documentclass[12pt,aps,prb,reprint,showpacs]{revtex4-1}

\usepackage{amsmath}    
\usepackage{graphicx}   
\usepackage{rotating}
\usepackage{lipsum}
 
\begin{document}

\title{Using graphical and pictorial representations to teach introductory astronomy students about the detection of extrasolar planets via gravitational microlensing}

\author{Colin S.\ Wallace}
\affiliation{Department of Physics and Astronomy, University of North Carolina at Chapel Hill, Chapel Hill, North Carolina 27599} 
\email{cswallace@email.unc.edu} 

\author{Timothy Chambers}
\affiliation{Department of Materials Science and Engineering, University of Michigan, Ann Arbor, Michigan 48109}
\email{timchamb@umich.edu}

\author{Edward E.\ Prather}
\affiliation{Center for Astronomy Education (CAE), Steward Observatory, University of Arizona, Tucson, Arizona 85721}
\email{eprather@as.arizona.edu} 

\author{Gina Brissenden}
\affiliation{Center for Astronomy Education (CAE), Steward Observatory, University of Arizona, Tucson, Arizona 85721}
\email{gbrissenden@as.arizona.edu}
\date{\today}

\begin{abstract}
The detection and study of extrasolar planets is an exciting and thriving field in modern astrophysics, and an increasingly popular topic in introductory astronomy courses.  One detection method relies on searching for stars whose light has been gravitationally microlensed by an extrasolar planet.  In order to facilitate instructors' abilities to bring this interesting mix of general relativity and extrasolar planet detection into the introductory astronomy classroom, we have developed a new \emph{Lecture-Tutorial}, ``Detecting Exoplanets with Gravitational Microlensing."  In this paper, we describe how this new \emph{Lecture-Tutorial}'s representations of astrophysical phenomena, which we selected and created based on theoretically motivated considerations of their pedagogical affordances, are used to help introductory astronomy students develop more expert-like reasoning abilities.
\end{abstract}

\maketitle

\section{Introduction}
\label{intro}

An incredible amount of our advances in understanding the universe over the past 100 years can be traced back to Einstein's groundbreaking 1915 paper on general relativity.  General relativity is used to understand the precession of Mercury's perihelion, the orbital decay of binary pulsars, and the expansion and evolution of the universe from the time of the Big Bang, among many other phenomena.  One of the most important uses of general relativity in observational astronomy is its ability to accurately describe how a massive object can act as a gravitational lens, bending and magnifying the light from a source located elsewhere in the universe.  While gravitational lensing is studied on the largest scales (e.g., galaxy clusters act as gravitational lenses for galaxies at higher redshifts), it also has observable effects at smaller scales, where gravitational lensing effects are used to detect the existence of extrasolar planets.\cite{perryman11}

This planet detection method relies on the fact that the light from a distant star may be lensed by objects in a closer stellar system.  Depending on the relative alignment of Earth, the distant star, and the closer stellar system, the light from the distant star may be lensed by a planet, star, or both in the stellar system.  In all cases, the angular separation between the multiple images that are typically formed by these lensing events are too small (on the order of 1 milliarcsecond) to be resolved by telescopes on Earth.  Hence, we call this lensing effect gravitational microlensing, in order to distinguish it from macrolensing events, in which the multiple images can be resolved.\cite{perryman11} Despite the fact that the gravitationally lensed images cannot be resolved, we can still determine that a microlensing event is taking place by observing the lensed star over an extended time period.  The motion of the lensing stellar system with respect to the distant star gradually changes the alignment between the Earth, stellar system, and distant star, which affects the amount of lensing that occurs.  Over tens of days, the light we detect from the distant star may appear to increase and then decrease in brightness as the star in the stellar system moves into and out of alignment with our line-of-sight to the distant star.  An extrasolar planet in the stellar system may cause a similar brightening and then dimming of the distant star, but on a time scale that may be as short as a few hours.\cite{perryman11, han07}  Using gravitational microlensing to detect extrasolar planets thus requires observations of a large patch of sky over an extended period of time.

Liebes (1964)\cite{liebes64} was the first to realize that gravitational lensing could be used to detect extrasolar planets, although the search for extrasolar planets using gravitational microlensing began with the work of Mao and Paczynski (1991).\cite{mao91}  The first confirmed detection of an extrasolar planet by gravitational microlensing was by Bond et al.\ (2004).\cite{bond04}  To date, thirty-seven planets have been discovered using this technique.\cite{exo15}  While this represents approximately only 2\% of all known extrasolar planets, gravitational microlensing provides the ability to detect Earth-mass planets around 1 AU from their parent stars.\cite{perryman11}  This mass and distance combination is interesting since such extrasolar planets could hypothetically harbor life as we know it.

The detection and study of extrasolar planets lies at the cutting-edge of modern astrophysics\cite{perryman14}, and it is also a topic that excites the imagination of general education students and the public at large.  Consequently, we developed new active engagement curricular materials for teaching this topic in college-level general education introductory astronomy courses (hereafter Astro 101). These materials, informed by decades of work into how people learn, especially in STEM disciplines\cite{bransford00, per13, freeman14, singer12, prather09a}, include new \emph{Lecture-Tutorials}, Think-Pair-Share activities, homework assignments, lecture slides, and assessment questions focused on the detection of extrasolar planets.  We created these active learning instructional activities and assessments to provide Astro 101 instructors with an easy-to-implement, evidence-based curriculum that meaningfully engages students in intellectually and scientifically robust collaborative discussions about extrasolar planets.  This article focuses on one of these new \emph{Lecture-Tutorials}, which addresses the detection of extrasolar planets via gravitational microlensing. 

While the design and development process for this new \emph{Lecture-Tutorial} mirrored the process we used in the development of previous \emph{Lecture-Tutorials}\cite{prather04, prather13, wallace12}, the topic of gravitational lensing posed an interesting challenge.  This topic is traditionally taught using representations (written, mathematical, graphical, etc.) that require a level of  understanding of physics and mathematics appropriate for advanced undergraduates or graduate students.  How could we design an activity that engages Astro 101 students, who do not possess the mathematical preparation necessary to interact with these representations, to do more than simply recall declarative knowledge?  While over 35\% of Astro 101 students in a national survey\cite{rudolph10} listed their major as being in science, engineering, or architecture, many of these students are at the beginning of their undergraduate careers and are years away from being able to take a course in general relativity.  The other 65\% of Astro 101 students are majors in the arts, humanities, social sciences, education, and professional fields (e.g., business, nursing, etc.)\cite{rudolph10}; instruction that is at too high or too low of a cognitive level may adversely affect their overall attitudes toward science, which has important ramifications given that they represent the nation's future journalists, historians, business leaders, teachers, politicians, parents, taxpayers, and voters.  In order to create an activity that inspires and engages Astro 101 students at an appropriate cognitive level, we had to judiciously develop a constellation of representations appropriate for the population of interest.  Each representation we created has its own pedagogical value and set of constraints.   When working in concert together, these representations can be very effective at elevating students' conceptual understandings and reasoning abilities related to the detection of extrasolar planets via gravitational microlensing.  Our ultimate goal for the new \emph{Lecture-Tutorial} is to enable Astro 101 students to engage in more expert-like reasoning about a general relativistic phenomenon and develop a greater discipline fluency.

In this article, we describe how we selected and designed the representations we used in the \emph{Lecture-Tutorial} ``Detecting Exoplanets with Gravitational Microlensing."  We first delve into the mathematics underlying gravitational microlensing and illustrate why we needed to emphasize certain aspects of our representations and deemphasize others in order to make it possible for students to develop physically correct understandings.  We then use the theoretical framework of Linder (2013)\cite{linder13} and Fredlund et al.\ (2014)\cite{fredlund14} to analyze the affordances of the different representations and to explain how they work together to aid in the development of students' discipline fluency.  In this paper, the ``affordances" of a representation are defined as the elements of disciplinary content that students are able to access and reason about using the representation.  This article provides a unique example of how to present advanced topics, such as general relativity, to a novice audience.  Additionally, this article details the difficult considerations that STEM curriculum developers must work through as they try to select and design appropriate new pedagogical representations that engage students at deep and meaningful levels.

\section{Dependence of brightness on the mass of the lensing object}
\label{brightness}

Before discussing the details of gravitational microlensing, we wish to introduce a key concept in observational astronomy: the light curve.  A light curve is a graph of the observed brightness of an astronomical body as a function of (in this case) time.  Light curves have many applications; for example, the decay rate of a supernova's light curve is an important factor in identifying the type of supernova.  As another example, periodic ``dips" in the light curve of a star can reveal the presence of planets transiting the star as they orbit, as studied by the recent Kepler mission.  Because of their broad applicability as well as their relative approachability for a novice, we suggest that the light curve is an excellent representation for teaching about a variety of astrophysical phenomena. 

When a single object, such as star, acts as a gravitational microlens for a distant star, we observe a time-symmetric light curve that shows how the image of the distant star brightens and then dims over the course of the microlensing event.  Microlensing causes the solid angle subtended by the star to change.  This change in solid angle affects the number of photons we receive from the star, leading to the observed change in its apparent brightness.  When the lensing system includes both a star and a planet, the planet may also microlens the distant star, creating a perturbation to the light curve; see Han (2007) for a classification of the different types of planetary light curve perturbations.\cite{han07}  

Every representation has its limitations, and those limitations may be explicit or implicit.  Instructors and curriculum designers must be aware of these limitations or else they run the risk of encouraging students to draw physically inappropriate conclusions from a given representation.  For example, one might be tempted to reason as follows: More massive planets will cause greater distortions in the geometry of spacetime, so one should be able to detect more photons coming to Earth and for a longer period of time than for cases with less massive planets.  This reasoning would then lead one to conclude that both the width and the height of the planetary perturbations in the light curves (such as those in Figures 6-9) are both correlated with the planet's mass.  Yet in the \emph{Lecture-Tutorial} on this topic, we use only the width of perturbations in the light curve to reason about the masses of the objects causing them.  To motivate this and other such pedagogical decisions, we must employ the mathematics of general relativity to describe how a single object lenses the light emitted by a more distant source.

The purpose of the following mathematical treatment is to unpack for the reader the disciplinary content knowledge relevant to a full understanding of the physics underlying the phenomenon of microlensing.  We have chosen to use the symbolic mathematical representations below because the readership of the \emph{American Journal of Physics} possesses sufficient mathematical sophistication to access their meaning.  This enables us to bring to light for the reader the origins of the quantitative relationships we are interested in teaching students to reason about in an expert-like manner.  We wish to emphasize that while the results derived here motivate the pedagogical choices we make as curriculum designers, we do not use the mathematical representation directly in our curriculum as Astro 101 students do not have the prerequisite knowledge for unpacking this representation. 

We used Hartle's text, \emph{Gravity: An Introduction to Einstein's General Relativity}, to inform our mathematical understanding of gravitational lensing.\cite{hartle03}  For the present discussion, we wish to derive the answer to the question:  How do lens mass and lens alignment affect the light curve of the distant light source?  The basic outline of this derivation is that we use the bending angle of light rays to determine the change in solid angle subtended by the image of a distant source when its light is lensed.  This change in the apparent area of the object will lead to a corresponding change in the apparent brightness.  The physical mechanism at work is that the gravitational field of an object between Earth and the distant source will curve the trajectories of photons, which then arrive at Earth from various angles.  This set of angles results in an image that is distorted from its original size and shape.  Note that we use gravitational units ($G = c = 1$), which are commonly used in general relativistic calculations.  This choice results in mass and length having the same units, allowing direct comparison between quantities of mass and distance.

Since we are interested in the detection of extrasolar planets by microlensing, we assume a pointlike lensing object of mass $M$. The deflection angle $\alpha$ for a light ray, which is the difference in angle between its initial and final trajectories, passing by the mass with an impact parameter $b >> M$ is
\begin{equation*}
\alpha = 2 R_s / b
\end{equation*}
where $R_s$ is the Schwarzschild radius of the lensing object.

Most of the bending of the light ray takes place over a distance of order $R_s$.  Because interstellar distances are vastly larger than this length scale, it is an excellent approximation to treat the trajectory of the light ray as being straight everywhere except at a single point, where the line will bend through the full angle $\alpha$.  This approximation is called the thin lens approximation.  We can now solve the problem using only Euclidean trigonometry, following the diagram shown in Figure 1.


In Figure 1, the angles are all extremely small, so $\sin x \approx \tan x \approx x$ for the angles $\alpha$, $\theta_S$, $\theta_I$.  These approximations give the horizontal distances labeled at the top of Figure 1, from which we write
\begin{equation*}
\theta_I D_S = \theta_S D_S + \alpha D_{LS}.
\end{equation*}
Next, we substitute $\alpha = 2 R_s /b$ and $b \approx \theta_I D_L$, then rearrange to find:
\begin{equation*}
\theta_I^2 - \theta_S \theta_I = \frac{2 R_s D_{LS}}{D_S D_L}.
\end{equation*}
We now have an expression for the positions $\theta_I$ at which the images of the source appear in the sky.  Both sides of this equation have units of radians squared, making it reasonable to define an angle $\theta_E$:
\begin{equation*}
\theta_E \equiv \left(2 R_S \frac{D_{LS}}{D_S D_L} \right)^{1/2}.
\end{equation*}
With this definition, we can solve for the two values of $\theta_I$ more easily and find that 
\begin{equation*}
\theta_\pm = \frac{1}{2} \left( \theta_S \pm \left( \theta_S^2 + 4 \theta_E^2 \right)^{1/2} \right).
\end{equation*}
The physical meaning of $\theta_E$ can be found by considering the case of perfect alignment between source, lens, and observer.  Substituting $\theta_S = 0$ into our solution yields $\theta = \theta_E$.  Thus, the angle $\theta_E$ is the angular radius of the ring of images formed in the case of perfect alignment (an Einstein ring).

Note that the radius of the Einstein ring is directly proportional to the square root of the mass of the lens: $\theta_E \propto \sqrt{M}$.  This will be important in answering the original question of how the shape of the microlensing light curve depends on the lens mass.

The observed brightness of the image of the light source depends only on its intrinsic brightness and the solid angle it subtends.  Therefore, the ratio of the brightness of the lensed image to the brightness of the unlensed image is just the ratio of the solid angles subtended by the images.  The intrinsic brightness of the object is not affected by the gravitational lensing, so it cancels out in this ratio.  This cancellation will be another important factor in our analysis of the light curve.  As such,
\begin{equation*}
\frac{I_\pm}{I_{*}} = \frac{\Delta \Omega_\pm}{\Delta \Omega_{*}}
\end{equation*}
where $I_\pm$ and $I_{*}$ are the lensed and unlensed brightnesses, and $\Delta \Omega_\pm$ and $\Delta \Omega_{*}$ are the solid angles subtended by the lensed and unlensed images, respectively.


Using the previous result for $\theta_\pm$we can calculate how the solid angle subtended by an image will change when it is lensed.  Figure 2 shows the original image of an object, and then the two lensed images, as seen from the observer's point of view.  Since the angles $\theta_S$ and $\theta_I$ are very small, we will again use the approximation $\sin x \approx x$ for these angles.  The formulas for solid angle then become
\begin{equation*}
\Delta \Omega_{*} = \theta_S \Delta \theta_S \Delta \phi
\end{equation*}
\begin{equation*}
\Delta \Omega_\pm = \theta_\pm \Delta \theta_\pm \Delta \phi .
\end{equation*}
We can substitute these expressions into the brightness ratio:
\begin{equation*}
\frac{I_\pm}{I_{*}} = \frac{\theta_\pm \Delta \theta_\pm \Delta \phi}{\theta_S \Delta \theta_S \Delta \phi}.
\end{equation*}
Notice that $\Delta \phi$ will cancel out -- the azimuthal angle subtended by the original image does not affect the ratio of the lensed and unlensed brightnesses.  In the limit of small lensing effects, this simplifies further:
\begin{equation*}
\frac{I_\pm}{I_{*}} = \left( \frac{\theta_\pm}{\theta_S} \right) \left( \frac{d \theta_\pm}{d \theta_S} \right).
\end{equation*}
The second term can be found by differentiating our solution for the angles $\theta_\pm$ at which the lensed images appear with respect to the source angle:
\begin{equation*}
\frac{d \theta_\pm}{d \theta_S} = \frac{1}{2} \left( 1 \pm \frac{\theta_S}{\left( \theta_S^2 + 4 \theta_E^2 \right)^{1/2}} \right).
\end{equation*}
The first term can also be expressed in terms of only $\theta_S$ and $\theta_E$ by substituting the expression for $\theta_\pm$.  Combining these equations and simplifying yields:
\begin{equation*}
\frac{I_\pm}{I_{*}} = \frac{1}{4} \left( \frac{\theta_S}{\left( \theta_S^2 + 4 \theta_E^2 \right)^{1/2}} + \frac{\left( \theta_S^2 + 4 \theta_E^2 \right)^{1/2}}{\theta_S} \pm 2 \right).
\end{equation*}
This shows that the outer image is brighter than the original, and the inner image is dimmer than the original.

In the case of microlensing, we cannot distinguish the lensed images from each other.  Instead, we simply detect the sum of the light from the images.  The ratio of the total microlensed brightness to the original brightness is therefore
\begin{equation*}
\frac{I_{tot}}{I_{*}} =\frac{I_+ + I_-}{I_{*}} = \frac{1}{2} \left( \frac{\theta_S}{\left( \theta_S^2 + 4 \theta_E^2 \right)^{1/2}} + \frac{\left( \theta_S^2 + 4 \theta_E^2 \right)^{1/2}}{\theta_S} \right).
\end{equation*}
Rewriting the two terms with a common denominator allows us to combine them into a single expression:
\begin{equation*}
\frac{I_{tot}}{I_{*}} = \frac{\theta_S^2 + 2 \theta_E^2}{\theta_S \left( \theta_S^2 + 4 \theta_E^2 \right)^{1/2}}.
\end{equation*}
One can instead write the equation in terms of a dimensionless parameter $u \equiv \theta_S / \theta_E$:
\begin{equation*}
\frac{I_{tot}}{I_{*}} = \frac{u^2 + 2}{u \sqrt{u^2 +4}}.
\end{equation*}

Now we address the original question:  How do lens mass and lens alignment affect the light curve of the distant light source?  We need to consider both the width and the height of the peak.  This can be done by making $\theta_S$ (or equivalently, $u$) a function of time in the above equations.  This function starts at a maximum, decreases to some minimum value, and then increases back to a maximum.  Recall that the angular radius of the Einstein ring is proportional to the square root of the lens's mass: $\theta_E \propto \sqrt{M}$.  

The width (or duration) of the peak increases as the mass of the lensing object increases.  To first order, the duration of the lensing event is the time for which the distant source is within an angular distance $\theta_E$ of the lens, as seen from Earth.  A more massive lens has a larger value for $\theta_E$, and so it will take a longer amount of time for the relative motion of the source and lens to traverse this distance, for a given velocity.  We therefore say that a more massive lens causes a wider peak.

The height of the peak -- that is, the ratio of the maximum brightness to the unlensed brightness -- also increases as the mass of the lensing object increases.  This occurs when the lens and source are most closely aligned when seen from Earth.  In terms of the variables in our equations, this is when $\theta_S (t) = \theta_{Smin}$ (or when $u(t)=u_{min}$).  Examining this in terms of either formulation should give the same result.  At this time,
\begin{equation*}
\frac{I_{max}}{I_{*}} = \frac{\theta^2_{Smin} + 2 \theta_E^2}{\theta_{Smin} \left( \theta^2_{Smin} + 4 \theta_E^2 \right)^{1/2}}.
\end{equation*}
Given this value for $\theta_{Smin}$, increasing the mass of the lens will increase $\theta_E$.  The Einstein angle is a quadratic term in the numerator, and we know that $\theta_E \propto \sqrt{M}$, so the numerator is a linear function of the lens mass.  In contrast, the denominator grows less-than-linearly with the lens mass, since the term inside the square root function is quadratic in $\theta_E$, and therefore linear in $M$.  As such, the value of the overall fraction should increase as $M$ increases if the other variables are held constant.  That is to say, all else being equal, more massive lenses should produce greater brightening effects.

The same conclusion can be reached in the dimensionless formulation.  For a detectable microlensing event, the distant source will be inside the Einstein ring of the lens at the time of minimum separation, so $u_{min} \le 1$.  Consider the dimensionless equation for the maximum brightness:
\begin{equation*}
\frac{I_{max}}{I_{*}} = \frac{u_{min}^2 + 2}{u_{min} \sqrt{u_{min}^2 +4}}.
\end{equation*}
As $u_{min}$ decreases from 1 to 0, this fraction grows monotonically toward infinity, demonstrating that more closely aligning the source and lens always increases the change in brightness.  Since $u=\theta_S/\theta_E$, and $\theta_E$ grows as the square root of the lens mass, increasing the mass of the lens has the effect of decreasing $u$ for a given value of $\theta_S$.  Qualitatively, one can imagine that adding mass to the lens would expand its Einstein ring, and that a source at a fixed location would therefore become relatively closer to being aligned with the center of the lens.

In either formulation of the brightness equation, we find that a greater lens mass causes a greater brightening than a smaller lens mass, assuming that the separation angle as a fraction of the Einstein angle ($\theta_S/\theta_E$, aka $u$) is the same for both masses.  Consequently, it is tempting to represent extrasolar planets of different masses with light curve perturbations that differ in both width and height.  However, if we are comparing the planetary perturbations on two different light curves, we cannot automatically assume that both perturbations correspond to the same value of $u$.  If we want students to use the height of the planetary perturbations to reason about the relative masses of extrasolar planets, then we have to include the caveat that the perturbations always correspond to the same ratio between the separation angle and the Einstein angle.  There is an incredible pedagogical risk associated with trying to bring this level of discipline knowledge to the population of students with which we are working.  Not only would this caveat intellectually over burden many Astro 101 students with more novel variables than they have the capacity to simultaneously reason about, it is also not representative of commonly occurring scenarios investigated by active researchers in extrasolar planet detection.  So rather than unnecessarily limit the physical situations we offer learners to only a limited set of geometric configurations which do not fully represent what is found in the universe, we purposefully chose to use only the width of a planetary perturbation, and not its height, to convey information about the extrasolar planet's mass in the \emph{Lecture-Tutorial}.

Note that while some \emph{Lecture-Tutorials} do require students to physically interpret mathematical equations, the ``Detecting Exoplanets with Gravitational Microlensing" \emph{Lecture-Tutorial} does not.  While the mathematical representation employed here has rich affordances for a mathematically expert population, Astro 101 students do not have the expertise needed to engage with this representation meaningfully.  Confronting these students with this representation would therefore not only be ineffective for promoting learning, but rather could lead to a belief that the topic is impenetrable to them and that there is no point in attempting to learn about it.  The purpose of the above derivation was to afford instructors a richer understanding of the effects of the physical variables on the microlensing processes, so that they can have a better understanding of how the relative positions and masses of lensing bodies affect observable astronomical outcomes.  We hope this understanding will enable readers to more effectively engage their learners in appropriate modes of cognition about discipline-specific ideas related to the detection of extrasolar planets via gravitational microlensing.
	
Having said this, we pose to the reader the same challenge we encountered when designing curriculum on the topic.  Having followed a mathematical route to deeply understanding gravitational microlensing, and knowing that Astro 101 students cannot follow the same pathway, how might one effectively teach students to engage in deep, expert-like reasoning about the relationships between astrophysical variables and observable effects in microlensing scenarios?  In the next section, we describe in detail the theoretical pedagogical framework we applied in addressing this challenge.  

\section{Affordances of different microlensing representations}
\label{affordances}

Linder (2013)\cite{linder13} uses the term ``disciplinary affordance" to refer to the potential of a representation for providing access to disciplinary knowledge.  Different representations (e.g., written words, equations, diagrams, graphs, etc.) have different disciplinary affordances, and no single representation by itself captures all aspects of the physical situation it is modeling.  However, multiple representations may work together to create a ``collective disciplinary affordance," providing a more holistic model of the physical situation and more potential access points for the development of disciplinary knowledge and fluency.  From this perspective, learning involves recognizing the disciplinary affordances of multiple complementary representations and coordinating the information provided by those representations to reason about novel physical situations.  Instructors must strive to create learning environments which enable students to better understand the affordances of multiple discipline representations.

Astrophysicists who use gravitational microlensing to detect extrasolar planets use a wide variety of representations in their own work.  For the purposes of this section, we will discuss three representations.  Two are now familiar to the reader: light curves, which constitute the fundamental observational evidence for microlensing events, and mathematical equations that astrophysicists use to relate information from these light curves to the physical parameters of the extrasolar planetary system.  The third is a reduced-dimensional pictorial representation of the curved spacetime that determines the paths of lensed light.  In this representation, the task of depicting the curvature of four-dimensional spacetime is replaced by depicting a two-dimensional  ``sheet" that is deformed and then projected onto the page.  Since reasoning about and depicting higher-dimensional structures is very difficult cognitively and artistically, this reduced-dimensional representation (hereafter RdR) is an efficacious proxy that allows for appropriate qualitative reasoning about the curvature of spacetime.  

These representations are powerful, but Fredlund et al. (2014)\cite{fredlund14} raise an important caveat: They have evolved through a process of ``rationalization," meaning they have become densely packed with knowledge that was developed over an extended period of time and through a series of preceding arguments.  The graphical, pictorial and mathematical representations used to describe gravitational microlensing constitute dense packages of information, including trigonometry, impact parameters, the physics of lenses, flux, and a conceptual model of gravitation as the curvature of spacetime, to name just a few.  This dense packaging of information presents a significant challenge to Astro 101 students who are attempting to interpret these representations.  Furthermore, students studying the detection of extrasolar planets via gravitational microlensing must reckon with the idea that we are detecting light that comes neither from the planet itself nor from the star it orbits, but rather from another, even more distant star.   All of these factors combined make the detection of extrasolar planets by gravitational microlensing an incredibly intellectually challenging topic for Astro 101 students to reason about.

The solution, according to Fredlund et al., is to help students ``unpack" these representations so that students may ``come to `see' the parts of intended meaning that are not directly discernible in the representation."  Note that unpacking a representation often involves utilizing or even creating simpler representations.  Fredlund et al.\ attribute the effectiveness of many PER-validated instructional techniques to the fact that they unpack and disambiguate the disciplinary affordances of many physics representations.\cite{fredlund14}

In practice, it is often necessary to set a foundation for students to unpack a complex representation by first teaching them simpler representations that afford them access to the content embedded in the more complex representation.  For example, the ``Motion in Two Dimensions" tutorial from the University of Washington's \emph{Tutorials in Introductory Physics}\cite{mcdermott08} ultimately guides students to unpack the mathematical representation of the centripetal acceleration vector.  However, the tutorial can only accomplish this by first getting students to represent kinematic quantities by drawing and combining vector arrows. This representation is both approachable to novices yet still rich enough to be used in discourse between experts.  In this example, vector arrows are a type of representation whose affordances are apparent to a large enough population of students that they can be productively used by novices to unpack more complex representations.  But it is not always the case that the set of existing representations typically used by experts within a discipline contains the complete set of representations needed to unpack a sophisticated discipline-specific representation for a novice.  Instructors and curriculum developers may find they have to incorporate novel representations to help students unpack more advanced representations in the process of developing more expert-like discipline fluency.  

In our work to make advanced astrophysics topics more accessible to the Astro 101 population, we have found it necessary to create many representations that would not typically be found in a textbook or used by two experts in the discipline who are engaged in discourse with one another on the topic.  Rather, these novel representations are especially tailored for teaching and learning about a discipline-specific topic, and so we call them ``pedagogical discipline representations" (PDRs).  In some cases PDRs are significantly simplified versions of representations experts might use in a colloquium talk when discussing a topic; in others they can be highly contextualized representations with unique features that purposefully unpack specific discipline concepts, engage novice learners' pre-existing mental models, and promote critical discourse.  

In the ``Detecting Exoplanets with Gravitational Microlensing" \emph{Lecture-Tutorial}, we use a mix of PDRs and representations that would be familiar to microlensing experts.  We carefully chose each representation so that its affordances would be accessible to students and so that it would help them improve their abilities to reason about microlensing and extrasolar planet detection. 

A typical implementation of this \emph{Lecture-Tutorial}, following the procedure detailed in Prather et al.\ (2004)\cite{prather04}, is as follows:  First, the instructor teaches a short lesson ($\sim$ 15 minutes) on the topic, after which students work in small peer groups on the \emph{Lecture-Tutorial}.  In completing the \emph{Lecture-Tutorial}, students collaboratively discuss and reach consensus on a series of carefully sequenced and scaffolded conceptual questions.  Students are able to seek help from the instructor and/or teaching assistants, who engage the student learning groups in Socratic dialogue.  Afterward, the instructor debriefs the activity by eliciting students' questions and engaging in a class-wide discussion of their answers.  Following the debrief, the instructor can check students' understanding by administering a series of formative assessment questions designed for the specific \emph{Lecture-Tutorial}. 

Figure 3 shows the first representation students encounter in the \emph{Lecture-Tutorial}.  The arrows represent the paths of some photons emanating from a distant star.  As Figure 3 shows, those paths may be altered by something in a hidden region of spacetime between the distant star and Earth.  This PDR affords students the idea that different regions of spacetime can alter the path light takes by different amounts.  From this representation, students can start to reason about how a distant star may appear brighter to an observer on Earth when more photons are re-directed (or lensed) toward Earth.  This establishes an important conceptual foundation students need in order to understand why the brightness of a distant star may change due to gravitational microlensing.  Additionally, it helps students begin to understand that the light we detect comes from the distant star, and not from the objects in the stellar system responsible for lensing that distant star's light.

\begin{figure}
\includegraphics[scale=0.55]{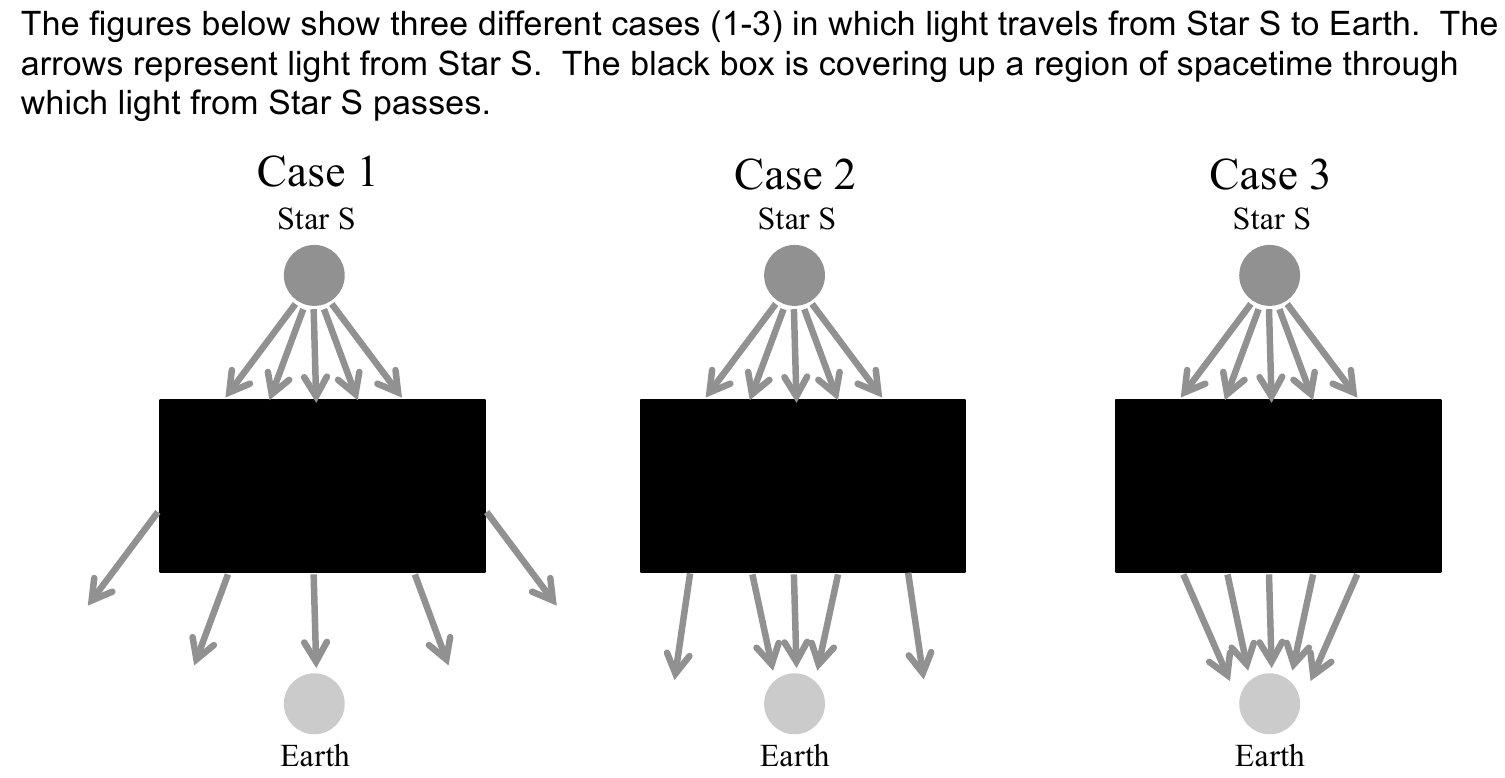}
\caption{\label{figure3}The first representation in the \emph{Lecture-Tutorial} shows students how a region of spacetime (hidden by the black boxes) may deflect light by different amounts.}
\end{figure}

We deliberately hid the objects in this region of spacetime with a black box in the Figure 3 PDR.  We could have included the representations of the lensing objects' masses and the corresponding curvature of spacetime in these regions, which would have enabled the PDR in Figure 3 to convey more information about the connection between mass, spacetime curvature, and the lensing of the light rays.  However, decades of research show that students are most likely to develop deep and long-lasting knowledge when they are actively engaged in the construction of that knowledge.\cite{bransford00, per13, freeman14, singer12, prather09a, prather04} Restricting the amount of information displayed in Figure 3 protects students from needing to simultaneously engage in two novel tasks, namely reasoning about the bending of light rays and processing the RdR.  This restriction therefore reduces the cognitive load experienced by students, enabling them to more effectively engage in constructing their own understanding about the relationship between the degree of bending of light rays and the amount of spacetime curvature (and, hence, the mass present) in a region of spacetime.  After developing this understanding, students can then apply their full cognitive resources to engaging with the \emph{Lecture-Tutorial}'s use of the RdR, as shown in Figure 4.  This figure presents three different RdRs of the curvature of spacetime for the region between the distant star and Earth, but not the paths taken by light emitted by the distant star.  Students are asked to match the cases shown in Figure 4 with those shown in Figure 3 and to rank the different representations of curved spacetime based on the amount of mass they must contain.

\begin{figure}
\includegraphics[scale=0.55]{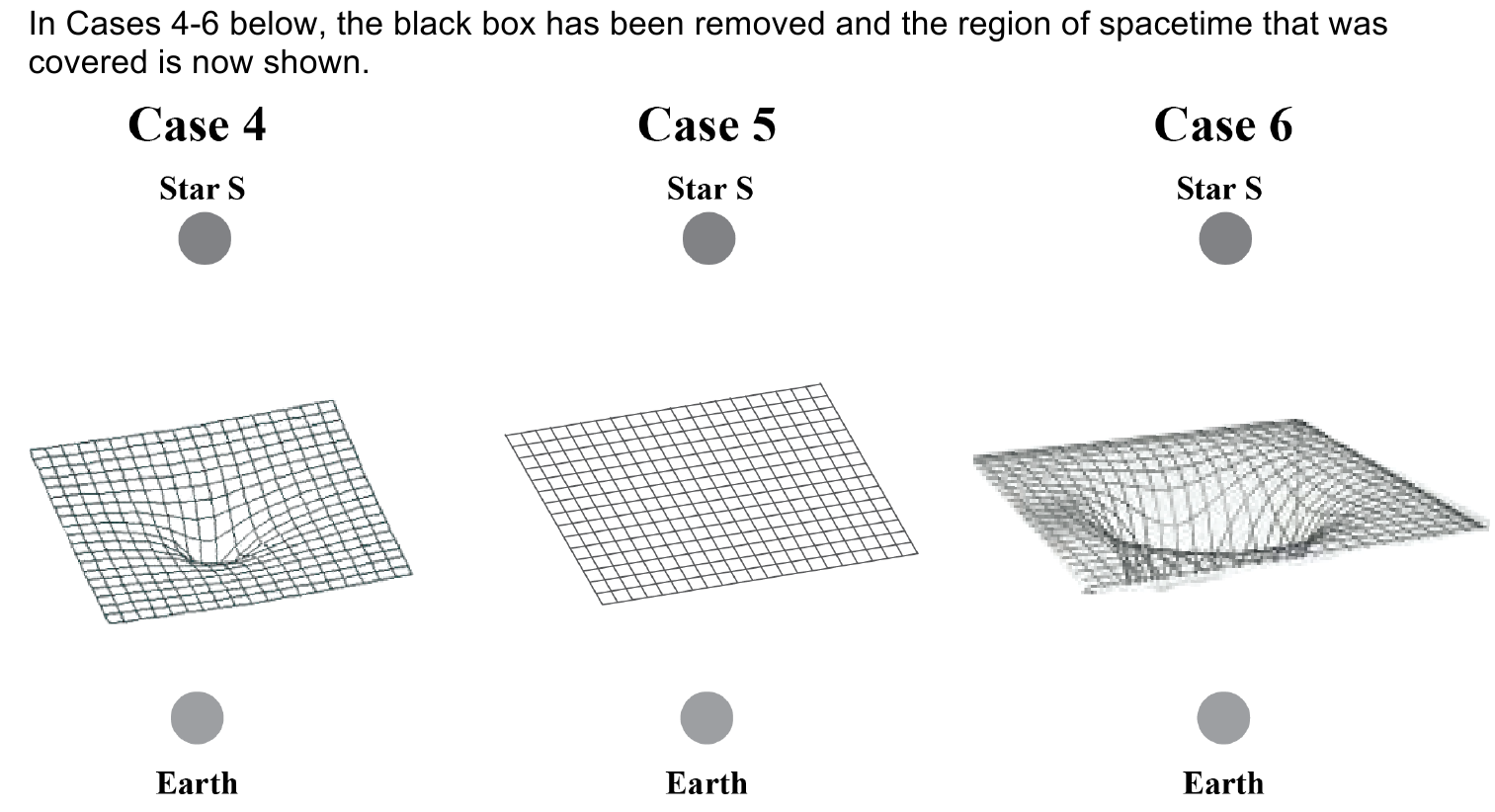}
\caption{\label{figure4}This reduced-dimensional representation (RdR) from the \emph{Lecture-Tutorial} shows the amount of curvature of the region of spacetime between the distant star and Earth.}
\end{figure}

Used together in this way, the representations provided in Figures 3 and 4 help students construct a mental model of the gravitational microlensing process for the case in which the lensing system comprises a single mass and when there is no relative motion between that mass, the distant star, and Earth.  In reality, gravitational microlensing events are transient phenomena and the lensing system may contain multiple masses.  Consequently, we use the representation in Figure 5 to introduce and illustrate the dynamic (as opposed to static) nature of a microlensing event in which both an extrasolar planet and its parent star lens the light of a distant star.  This representation provides students the opportunity to begin to reason about which object (the star or planet) creates a larger warping of spacetime.  It also provides students key information they will need in order to determine which of the light curves shown in Figure 6 would result from this particular arrangement of planet and parent star.  The \emph{Lecture-Tutorial} is designed to help students reason as follows: Since the planet will move through the region of spacetime between the distant star and Earth before the parent star, the increase in brightness of the distant star must occur at an earlier time than the brightening of the distant star resulting from the lensing caused by the parent star.  Using these PDRs, the \emph{Lecture-Tutorial} guides students to conclude that Graph D in Figure 6 corresponds to the situation shown in Figure 5.  Creating and understanding a complete, correct mapping between these two representations is a critical step in a student's development of a mental model of the microlensing process.

\begin{figure}
\includegraphics[scale=0.6]{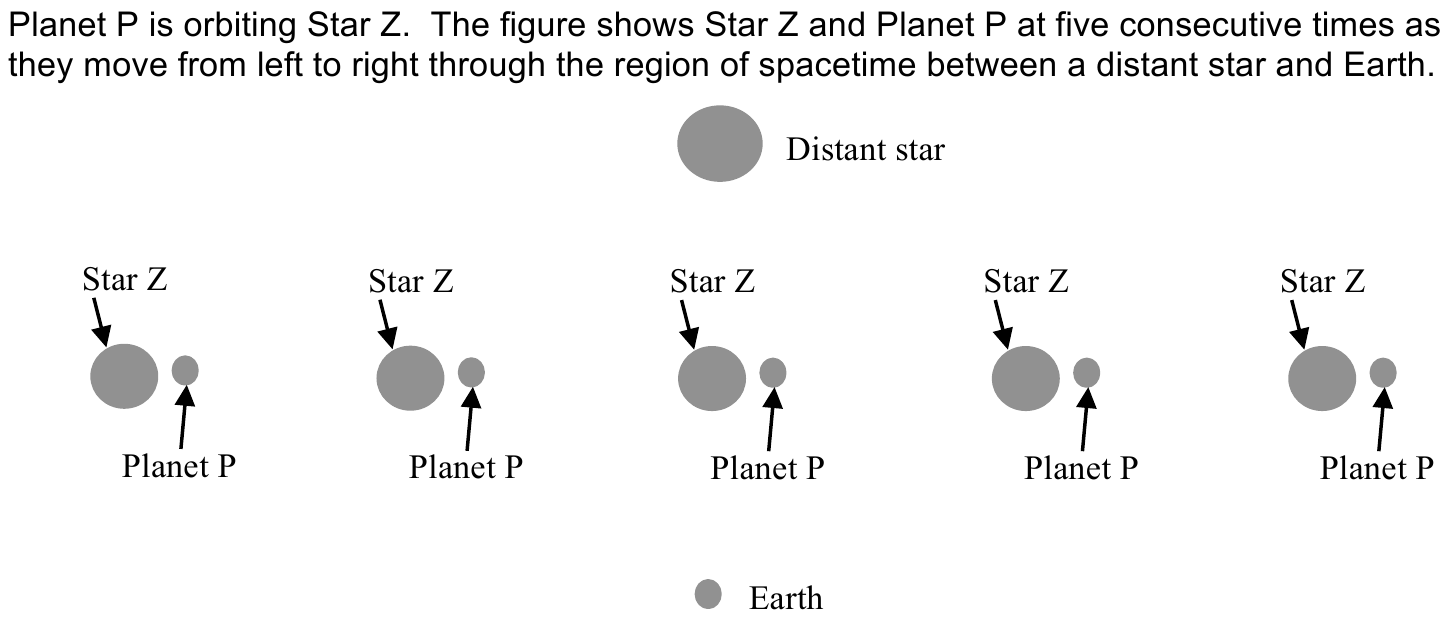}
\caption{\label{figure5}This representation from the \emph{Lecture-Tutorial} helps students visualize the movement of an extrasolar planet and its parent star relative to a distant star, as seen from Earth.}
\end{figure}

\begin{figure}
\includegraphics[scale=0.57]{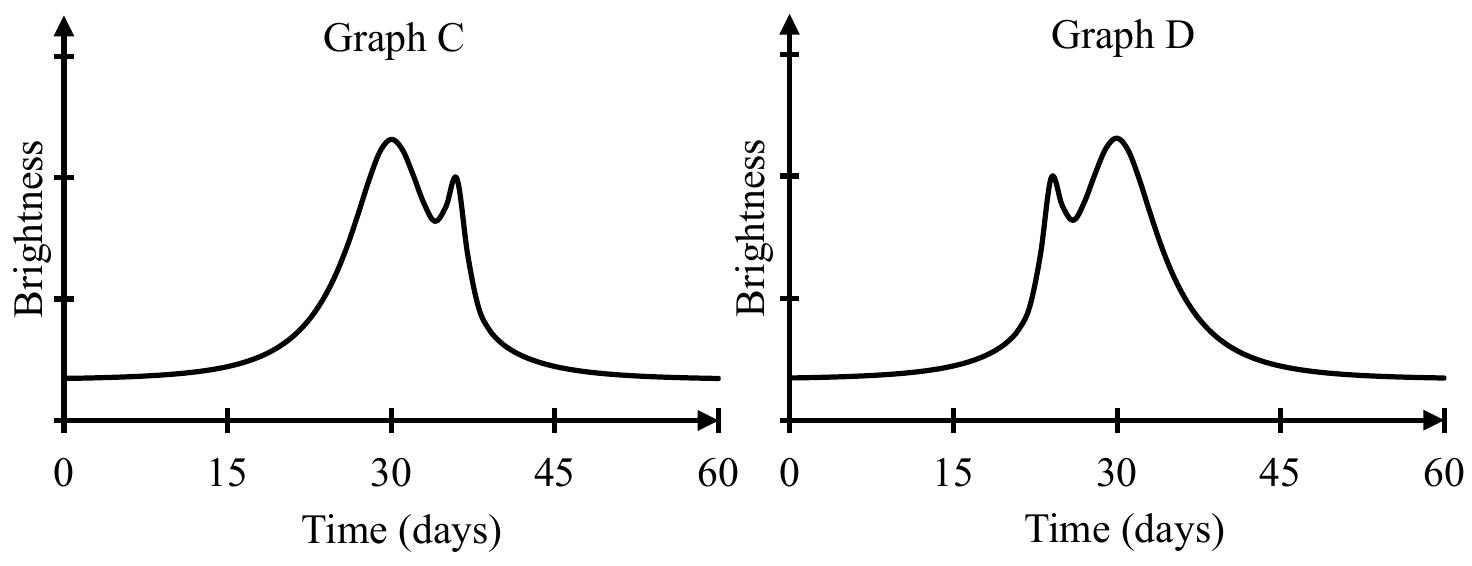}
\caption{\label{figure6}Students must determine which of these two light curves corresponds to the situation depicted in Fig. 5.}
\end{figure}

Figure 7 shows four different light curves.   At this point in the \emph{Lecture-Tutorial}, students have progressed through a carefully sequenced set of questions that guide their thinking to investigate and interpret the PDRs in Figures 3-6.  The tasks presented thus far provide students with the foundational discipline knowledge necessary to recognize that each of the light curves in Figure 7 must be due to a system of one extrasolar planet and its parent star, which each lens the light of a distant star.  They are asked to rank the relative masses of the extrasolar planets based on the widths of the perturbations in the corresponding light curves of Figure 7.  They are also asked to determine in which cases the extrasolar planet was located to the left or right of its parent star, as seen from Earth, during the microlensing event.

\begin{figure}
\includegraphics[scale=0.55]{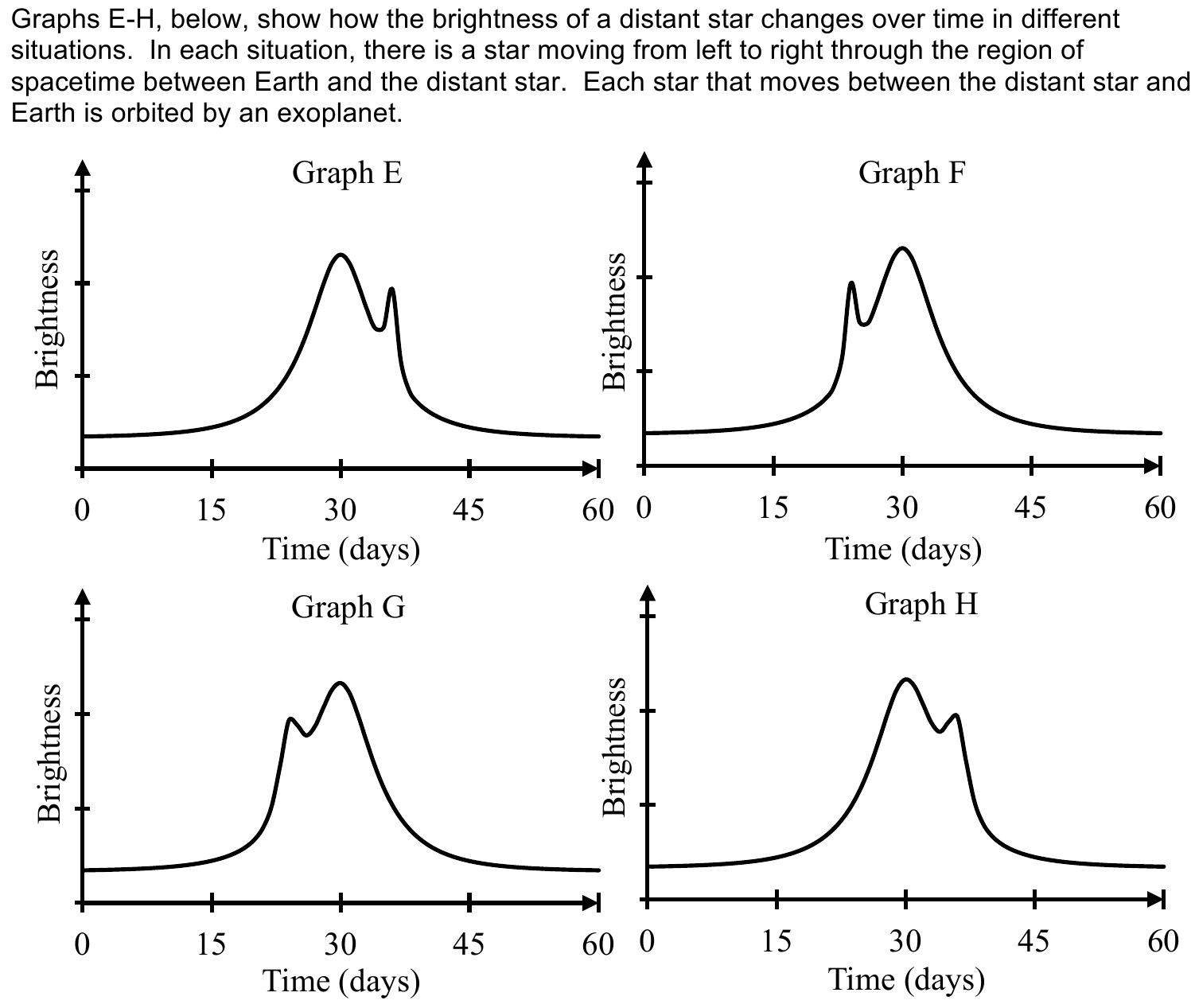}
\caption{\label{figure7}Four light curves that students reason about in terms of the mass of the detected extrasolar planet and whether or not it is located to the left or right of its parent star, as seen from Earth.}
\end{figure}

Figure 8 shows two additional light curves.  These curves differ in the temporal separation of their planetary perturbations.  This affords students the opportunity to reason about which light curve must be due to the stellar system in which the extrasolar planet was at the greater physical distance from its parent star, as seen from Earth.

\begin{figure}
\includegraphics[scale=0.55]{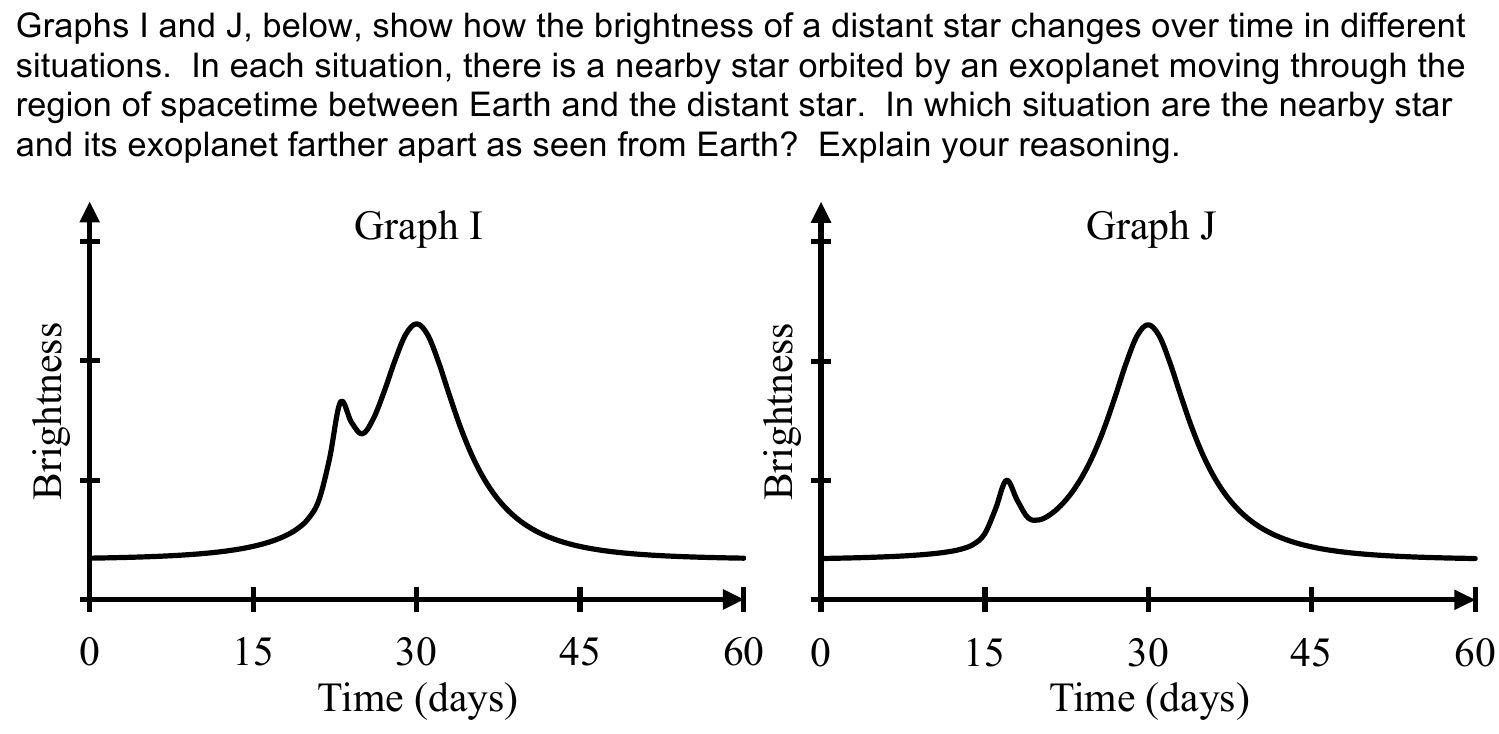}
\caption{\label{figure8}Two light curves students reason about to determine which of the detected extrasolar planets has the greatest separation from its parent star.}
\end{figure}

The final question of the \emph{Lecture-Tutorial} asks students to reason about the stellar system that must have been responsible for producing the light curve shown in Figure 9.  This question probes whether students have a complete and coherent mental model for the full breadth of physical characteristics of the extrasolar planet system that this light curve represents.  After working through all preceding components of the \emph{Lecture-Tutorial}, students should be able to discern how the features of this representation affords information about the number of extrasolar planets in the stellar system, the relative separation of those planets from their parent star, and the relative masses of those planets.  Figure 10 shows an analogous question that we use as part of a post-\emph{Lecture-Tutorial} homework assignment; in this question, students must sketch a light curve based on the stellar system that is shown.  Our post-instruction assessments of student understanding show that most of the Astro 101 students participating in this work -- the majority of whom are non-STEM majors\cite{rudolph10} -- are able to successfully complete the reasoning tasks shown in Figures 9 and 10.

\begin{figure}
\includegraphics[scale=0.55]{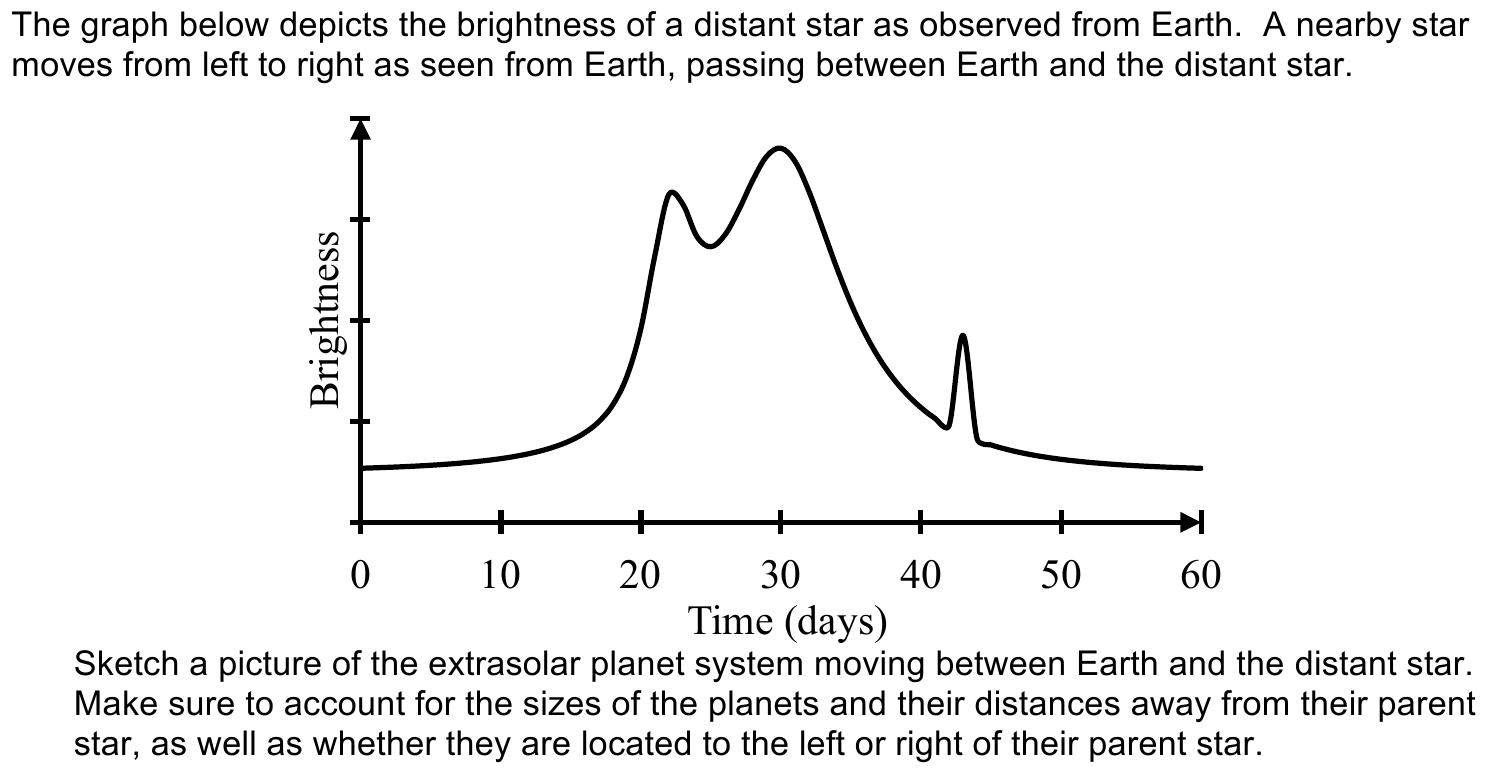}
\caption{\label{figure9}In the final question of the \emph{Lecture-Tutorial}, students must reason about the number of extrasolar planets in the stellar system, the relative separation of those planets from their parent star, and the relative masses of those planets.}
\end{figure}

\begin{figure}
\includegraphics[scale=0.55]{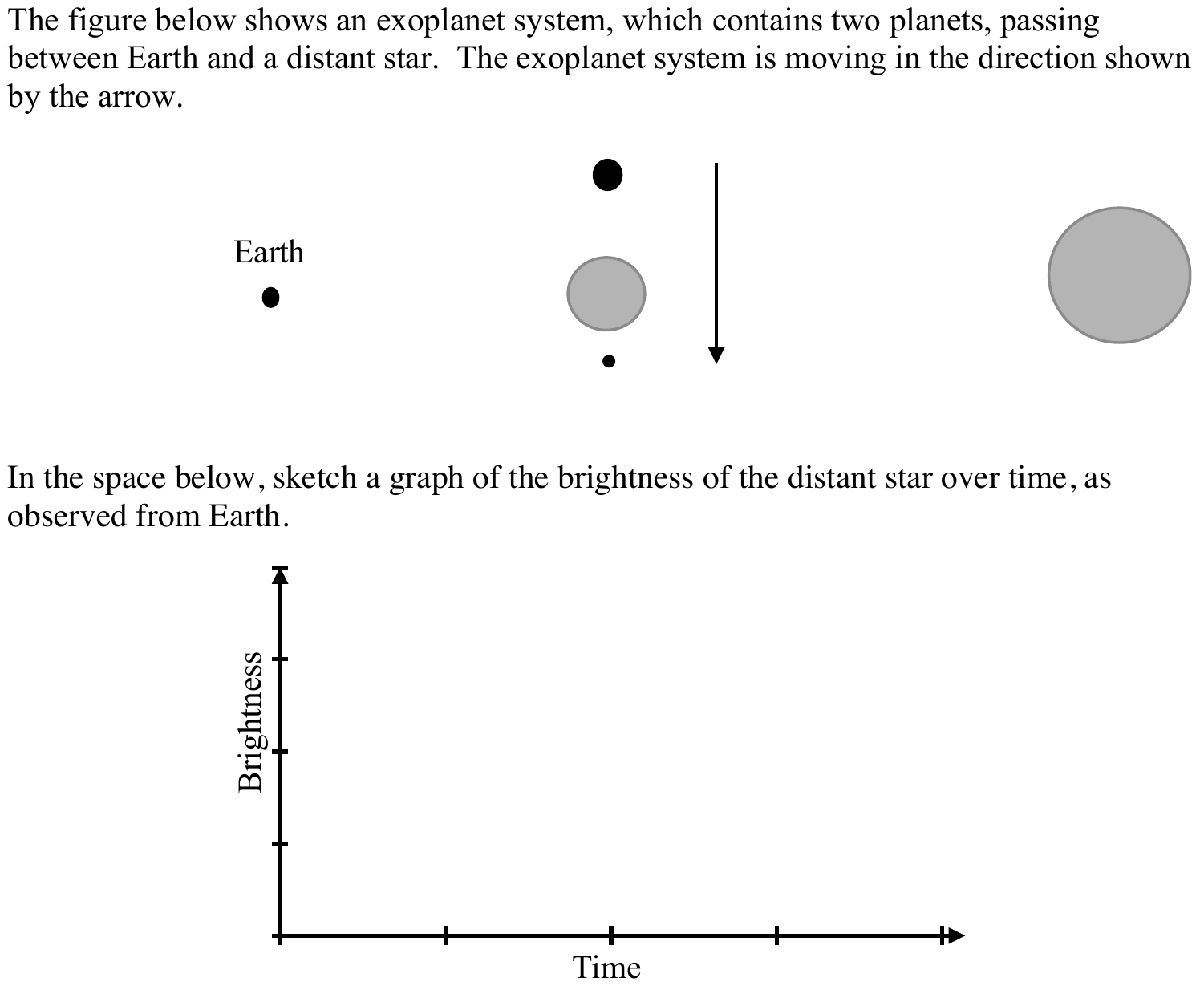}
\caption{\label{figure10}A question from the \emph{Lecture-Tutorial}'s associated homework that asks students to sketch the light curve that the stellar system would produce when it lenses the light of a distant star.}
\end{figure}

While our assessment of students' post-instruction understandings of this topic are ongoing, we have found that between 70-85\% of students can correctly answer multiple-choice midterm and final exam questions that are modeled after the \emph{Lecture-Tutorial} and homework questions in Figures 9 and 10.  These values are consistent with the results from multiple studies conducted over the past fifteen years on how student understanding of difficult-to-understand astrophysics topics is significantly elevated above what is achieved in lecture-based classrooms when they engage with research-based active learning curricula.\cite{prather09a, prather04, wallace12, hudgins06, prather09b, williamson13}  We attribute the greater discipline fluency that these students attain to their cognitive engagement with the \emph{Lecture-Tutorial}'s carefully targeted sequence of Socratic-style questions, which guide students to unpack the content afforded to them by the PDRs used in Figures 3-9.  

\section{Conclusions}
\label{conclusions}

Gravitational microlensing is an important application of Einstein's general theory of relativity to extrasolar planet detection, which is at the forefront of modern astrophysics.  We have developed a new \emph{Lecture-Tutorial}, ``Detecting Exoplanets with Gravitational Microlensing," that instructors can use to actively engage their Astro 101 students.  This paper details the various representations the \emph{Lecture-Tutorial} employs to help students develop their abilities to reason about this topic.  While some of these representations are familiar to experts in the field of extrasolar planet detection (e.g., light curves), others are novel representations that we created specifically for the purposes of teaching and learning, and which we refer to as ``pedagogical discipline representations" (PDRs).  We used the theoretical framework of Linder (2013)\cite{linder13} and Fredlund et al.\ (2014)\cite{fredlund14} to analyze the affordances of the \emph{Lecture-Tutorial}'s representations and to justify why we selected this particular set and sequence of representations.  We also discussed the limitations of some representations, which underscores the importance of instructors understanding which conclusions can and cannot be drawn from a given representation.  One component of an expert instructor's pedagogical content knowledge\cite{shulman86} is a knowledge of the affordances and limitations of different PDRs and other representations; such knowledge enables an instructor to select the most pedagogically appropriate representations to help students efficiently develop greater discipline fluency with a given topic.  We hope that by illustrating our use of this theoretical framework for choosing and developing representations, other instructors and curriculum developers will be able to take complex, advanced topics, such as gravitational microlensing, and translate them into curricula that meaningfully engage and elevate the reasoning abilities of introductory students.

\begin{acknowledgments}
The development of the ``Detecting Exoplanets with Gravitational Microlensing" Lecture-Tutorial and accompanying curricula was funded by the generous contributions of NASA's Exoplanet Exploration Program.
\end{acknowledgments}

\end{document}